\documentclass{article}   

\usepackage{graphicx}
\usepackage[english]{babel}
\usepackage[utf8]{inputenc}
\usepackage{amsmath}
\usepackage{amssymb}
\usepackage{amsthm} 
\usepackage{hyperref}
\usepackage{tikz} 
\usepackage{multirow}
\usepackage{subcaption}
\usepackage{natbib}

\newtheorem*{algorithm*}{Algorithm}

\newenvironment{noname}
{\par\medskip\begin{itshape}}
{\end{itshape}\par\medskip}

\title{Approximate Maximum Likelihood Estimation}
\author{Johanna Bertl\footnote{johanna.bertl@clin.au.dk}\\
    {\normalsize Aarhus University, Denmark}
    \and
    Gregory Ewing \\
    {\normalsize \'Ecole polytechnique f\'ed\'erale de Lausanne, Switzerland}
    \and
    Carolin Kosiol \\
	{\normalsize Vetmeduni Vienna, Austria}
    \and
    Andreas Futschik \\
	{\normalsize Johannes Kepler University Linz, Austria}
}

\begin{document}
\maketitle

\begin{abstract}
	In recent years, methods of approximate parameter estimation have attracted considerable interest in complex problems where exact likelihoods are hard to obtain. In their most basic form, Bayesian methods such as Approximate Bayesian Computation (ABC) involve sampling from the parameter space and keeping those parameters that produce data that fit sufficiently well to the actually observed data. Exploring the whole parameter space, however, makes this approach inefficient in high dimensional problems. This led to the proposal of more sophisticated iterative methods of inference such as particle filters.
	
	Here, we propose an alternative approach that is based on stochastic gradient methods and applicable both in a frequentist and a Bayesian setting. By moving along a simulated gradient, the algorithm produces a sequence of estimates that will eventually converge either to the maximum likelihood estimate or to the maximum of the posterior distribution, in each case under a set of observed summary statistics. To avoid reaching only a local maximum, we propose to run the algorithm from a set of random starting values. 
	
	As good tuning of the algorithm is important, we explored several tuning strategies, and propose a set of guidelines that worked best in our simulations. We investigate the performance of our approach in simulation studies, and also apply the algorithm to two models with intractable likelihood functions. First, we present an application to inference in the context of queuing systems. We also re-analyze population genetic data and estimate parameters describing the demographic history of Sumatran and Bornean orang-utan populations. 

\end{abstract}

\section{Introduction}

Both in the Bayesian as well as in the frequentist framework, statistical inference commonly uses the likelihood function. However, under a wide range of complex models, no explicit formula is available. This occurs for example when modelling population genetic and evolutionary processes \citep{MarjoramTavare06}, in spatial statistics \citep{SoubeyrandEtAl09}, and in queuing systems \citep{HegglandFrigessi04}. Furthermore, such situations also occur with dynamical systems as used for example in systems biology \citep{ToniEtAl09}, and epidemiology \citep{McKinleyEtAl09}. Here, we consider statistical models with an intractable distribution theory, but a known data generating process under which simulations can be obtained.

A recent approach to overcome this problem is Approximate Bayesian Computation (ABC) \citep{BeaumontEtAl02}. Its basis is a rejection algorithm: Parameter values are randomly drawn from the prior distribution, data sets are then simulated under these values. To reduce complexity, informative but low dimensional summaries are derived from the data sets. All parameter values that gave rise to summary statistics similar to those computed for the observed data are then accepted as a sample from the posterior distribution. Hence, ABC does not involve any kind of direct approximation of the likelihood, but with a uniform prior (with support on a compact subset of the parameter space taken large enough to contain the maximum likelihood estimator) it can be used to obtain a simulation-based approximation of the likelihood surface and, subsequently, the maximum likelihood estimate for the observed summary statistic values \citep{CreelKristensen13, RubioJohansen13}.

ABC has successfully been used in population genetic applications and also in other fields \citep{Beaumont10}. However, the sampling scheme can be inefficient in high-dimensional problems: The higher the dimension of the parameter space and the summary statistics, the lower is the acceptance probability of a simulated data set. Consequently, more data sets need to be simulated to achieve a good estimate of the posterior distribution. Alternatively, the acceptance threshold needs to be relaxed, but this increases the bias in the approximation to the posterior distribution. To reduce the sampling space, combinations of ABC and iterative Monte Carlo methods such as particle filters and Markov Chain Monte Carlo have been proposed \citep{BeaumontEtAl09, WegmannEtAl09}.

Here, we follow an alternative approach to obtain an approximate maximum likelihood estimate. Instead of using random samples from the whole parameter space, we propose a stochastic gradient algorithm that approximates the maximum likelihood estimate. Similar to ABC, it relies on lower-dimensional summary statistics of the data. In the simultaneous perturbations algorithm, in each iteration two noisy evaluations of the likelihood are sufficient to obtain an ascent direction \citep{Spall92}. To this end the likelihood is approximated by kernel density estimation on summary statistics of data simulated under the parameter value of interest.

Our algorithm is related to a method suggested in \citet{DiggleGratton84} \citep[see also][]{FermanianSalanie04}.  There, an approximate maximum likelihood estimate is obtained using a stochastic version of the Nelder-Mead algorithm. However, the authors explore only applications to 1-dimensional \textit{i.i.d.} data. In principle, our approach can also be applied in the context of indirect inference, where summary statistics are derived from a tractable auxiliary model \citep{DrovandiEtAl11}.

\section{Method}

\label{sec:method}

We will start by describing the so called simultaneous perturbation algorithm for optimizing the expected value of a random function using simulations. Next, we explain how this algorithm can be adapted to obtain maximum likelihood estimates. Details on the tuning can be found in the following section.

\subsection{General simultaneous perturbation algorithm}

Let $Y \in \mathbb{R}$ be a random variable depending on $\Theta \in \mathbb{R}^p$. The function $L(\Theta)=E(Y \mid \Theta)$ shall be maximized in $\Theta$, but $L(\Theta)$ as well as the gradient $\nabla L(\Theta)$ are unknown. If a realization $y(\Theta)$ of $Y \mid \Theta$ can be observed for any value of $\Theta$, the maximizer $\arg \max_{\Theta \in \mathbb{R}^p} L(\Theta)$ can be approximated by the simultaneous perturbation algorithm \citep{Spall92}. 

Similar to a deterministic gradient algorithm, it is based on the recursion
\begin{equation*}
	\Theta_k = \Theta_{k-1} + a_k \nabla L(\Theta_{k-1}),
\end{equation*}
where $a_k \in \mathbb{R}^+$ is a decreasing sequence. However, as $\nabla L(\Theta_{k-1})$ is unknown, it is substituted by a rescaled putative ascent direction that is randomly chosen among the vertices of the unit hypercube. Thus, in iteration $k$, a random vector $\delta_k$ with elements 
\begin{equation}
\label{eq:delta}
	\delta_k^{(i)} = \left\{ 
		\begin{array}{ll}
			-1 & \textrm{with probability } 1/2, \\
			+1 & \textrm{with probability } 1/2,
		\end{array}
	\right. 
\end{equation}
for $i = 1, \ldots, p$ is generated, and the ascent direction is approximated by 
\begin{equation*}
	\delta_k \frac{y\left(\Theta_k + c_k \delta_k \right) - y\left(\Theta_k - c_k \delta_k \right)}{2c_k},
\end{equation*}
with $c_k \in \mathbb{R}^+$ being a decreasing sequence.

\subsection{Approximate maximum likelihood algorithm}

Suppose, data $D_{obs}$ are observed under model $\mathcal{M}$ with unknown parameter vector $\Theta \in \mathbb{R}^p$. Let $L(\Theta; D_{obs}) = p(D_{obs} \mid \Theta)$ denote the likelihood of $\Theta$. For complex models often there is no closed form expression for the likelihood, and for high dimensional data sets, the likelihood can be  difficult to estimate. As in ABC, we therefore consider $L(\Theta; S_{obs}) = p(S_{obs} \mid \Theta)$, an approximation to the original likelihood that uses  a $d$-dimensional vector of summary statistics $S_{obs}$ instead of the original data $D_{obs}$. 

An estimate $\hat{L}(\Theta; S_{obs})$ of $L(\Theta; S_{obs})$ can be obtained from simulations under $\mathcal{M}(\Theta)$ using kernel density estimation. By setting $y(\Theta) = \hat{L}(\Theta; S_{obs})$, we adopt the simultaneous perturbation algorithm to approximate the maximum likelihood estimate $\hat{\Theta}_{\text{ML}}$ of $\Theta$. In more detail, the algorithm is as follows:
\medskip

\begin{algorithm*}[AML]\label{alg:AML}
	Let $a_k, c_k \in \mathbb{R}^+$ be two decreasing sequences. Let $H_k$ be a sequence of symmetric positive definite $d \times d$ matrices and $\kappa$ a d-dimensional kernel function satisfying $\int_{\mathbb{R}^d}{\kappa(x) dx}=1$. Choose a starting value $\Theta_0 \in \mathbb{R}^p$. 

\noindent For $k = 1, 2, \ldots, K$:
	\begin{enumerate}
		\item Choice of a likely ascent direction in $\Theta_{k-1}$:
			\begin{enumerate}

				\item\label{it:randomdir} Generate a random vector $\delta_k$ as defined in equation \eqref{eq:delta}.

				\item\label{it:simulate} Simulate datasets $D^-_1, \ldots, D^-_n$ from $\mathcal{M}(\Theta^-)$ and $D^+_1, \ldots, D^+_n$ from $\mathcal{M}(\Theta^+)$ with $\Theta^\pm = \Theta_{k-1} \pm c_k \delta_k$.

				\item\label{it:sumstat} Compute summary statistics $S^-_j$ on dataset $D^-_j$ and $S^+_j$ on $D^+_j$ for $j = 1, \ldots, n$.

				\item\label{it:kde} Estimate the likelihood $\hat{L}(\Theta^-;S_{obs})=\hat{p}(S_{obs} \mid \Theta^-)$ and $\hat{L}(\Theta^+;S_{obs})=\hat{p}(S_{obs} \mid \Theta^+)$ from the  summary statistics $S^-_1, \ldots, S^-_n$ and $S^+_1, \ldots, S^+_n$, respectively, with multivariate kernel density estimation \citep{WandJones}:
					\begin{equation*}
						\hat{L}(\Theta^-;S_{obs}) = \frac{1}{n \sqrt{\det(H_k)}} \sum_{j = 1}^{n}{\kappa \left( H_k^{-1/2}\left(S_{obs} - S_j^-\right)\right)}
					\end{equation*}
				and analogously for $\hat{L}(\Theta^+, S_{obs})$.

				\item\label{it:gradient} Estimate the ascent direction $\nabla l(\Theta_{k-1}; S_{obs})$ by
					\begin{equation*}
					 \hat{\nabla}_{c_k} \hat{l}(\Theta_{k-1}; S_{obs})  
					= \delta_k \frac{\log \hat{L}(\Theta^+; S_{obs}) - \log \hat{L}(\Theta^-; S_{obs})}{2c_k} 
					\end{equation*}
		\end{enumerate}

		\item Updating $\Theta_k$:
			\begin{equation*}
				\Theta_{k} = \Theta_{k-1} + a_k  \hat{\nabla}_{c_k} \hat{l}(\Theta_{k-1}; S_{obs})
			\end{equation*}

	\end{enumerate}

\end{algorithm*}

Then, the approximate maximum likelihood (AML) estimate is $\hat{\Theta}_{\text{AML}}:= \Theta_K$.

If, in a Bayesian setting, a prior distribution $\pi(\Theta)$ has been specified, the algorithm can be modified to approximate the maximum of the posterior distribution, $\hat{\Theta}_{\text{MAP}}$, by multiplying $\hat{L}(\Theta^-, S_{obs})$ and $\hat{L}(\Theta^+, S_{obs})$ by $\pi(\Theta^-)$ and $\pi(\Theta^+)$, respectively.

\subsection{Parametric bootstrap}

Confidence intervals and estimates of the bias and standard error of $\hat{\Theta}_{\text{AML}}$ can be obtained by parametric bootstrap: $B$ bootstrap datasets are simulated from the model $\mathcal{M}(\hat{\Theta}_{\text{AML}})$ and the AML algorithm is run on each dataset to obtain the bootstrap estimates $\hat{\Theta}_{\text{AML}, 1}^\ast,$ $ \ldots, \hat{\Theta}_{\text{AML}, B}^\ast$. This sample reflects both the error of the maximum likelihood estimator as well as the approximation error.

We compute simple bootstrap confidence intervals that are based on the assumption that the distribution of $\hat{\Theta}_{\text{AML}} - \Theta$ can be approximated sufficiently well by the distribution of $\hat{\Theta}_{\text{AML}}^\ast - \hat{\Theta}_{\text{AML}}$, where $\hat{\Theta}_{\text{AML}}^\ast$ is the bootstrap estimator. Then, a two-sided $(1-\alpha)$-confidence interval is defined as
\begin{equation*}
	\left[2\hat{\Theta}_\text{AML} - q_{(1-\alpha/2)}(\hat{\Theta}_\text{AML}^\ast), 2 \hat{\Theta}_\text{AML} - q_{(\alpha/2)}(\hat{\Theta}_\text{AML}^\ast) \right],
\end{equation*}
where $q_{(\beta)}(\hat{\Theta}_\text{AML}^\ast)$ denotes the $\beta$ quantile of $\hat{\Theta}^\ast_{\text{AML},1}, \ldots,$ $\hat{\Theta}^\ast_{\text{AML}, B}$ \citep{DavisonHinkley}.

\section{Tuning guidelines}

The performance of the algorithm strongly depends on the choice of the sequences $a_k$ and $c_k$. However, their optimal values depend on the unknown likelihood function. Another challenge in the practical application is the stochasticity: large steps can by chance lead very far away from the maximum.

\subsection{Choice of $a_k$ and $c_k$ and convergence diagnostics}

Here, we consider sequences of the form
\begin{equation*}
		a_k = \frac{a}{\left( k+A \right)^\alpha} \textrm{ and } c_k = \frac{c}{k^\gamma}
\end{equation*} 
as proposed in \citet{Spall}. To reflect the varying slope of $L$ in different directions of the parameter space, we choose $a \in \mathbb{R}^p$ (this is equivalent to scaling the space accordingly). Choosing $\alpha=1$ and $\gamma=1/6$ ensures optimal convergence speed \citep{Spall}. The optimal choice of $a$ and $c$ depends on the unknown shape of $L$. Therefore, we propose the following heuristic based on suggestions in \cite{Spall} and our own experience to determine these values as well as $A \in \mathbb{N}$, a shift parameter to avoid too fast decay of the step size in the first iterations. 

\begin{noname}
Let $K$ be the number of planned iterations and $b \in \mathbb{R}^p$ a vector that gives the desired stepsize in early iterations in each dimension. Choose a starting value $\Theta_0$. 

	\begin{enumerate}

		\item \label{it:c} Set $c$ to a small percentage of the total parameter range (usually between 1\% and 5\%) to obtain $c_1$. 

		\item Set $A = \lfloor 0.1*K \rfloor$.

		\item \label{it:a} Choose $a$: 

			\begin{enumerate}
				\item Estimate $\nabla l(\Theta_0; S_{obs})$ by the median of $n_1$ finite differences approximations (step 1 of the AML algorithm), $\bar{\hat{\nabla}}_{c_1} \hat{l}(\Theta_0, S_{obs})$. 

				\item Set
					\begin{equation*}
						a^{(i)} = \frac{b^{(i)} (A+1)^{\alpha}}{\left(\bar{\hat{\nabla}}_{c_1} \hat{l}(\Theta_0; S_{obs})\right)^{(i)}} \text{ for } i = 1, \ldots, p.
					\end{equation*}
			\end{enumerate}
	\end{enumerate}
\end{noname}

As $a$ is determined using information about the likelihood in $\Theta_0$ only, it might not be adequate in other regions of the parameter space. To be able to distinguish convergence from a too small step size, we simultaneously monitor the growth of the likelihood function and the trend in the parameter estimates to adjust $a$ if necessary. Every $K_0$ iterations the following three tests are conducted on the preceding $K_0$ iterates:

\begin{noname}
\begin{itemize}
	\item \textbf{Trend test (too small $a$):} For each dimension $i = 1, \ldots, p$ a trend in $\Theta^{(i)}_k$ is tested using the standard random walk model
	\begin{equation*}
		\Theta_{k}^{(i)} = \Theta_{k-1}^{(i)} + \beta + \epsilon_k,
	\end{equation*}
	where $\beta$ denotes the trend and $\epsilon_k \sim N(0, \sigma^2)$. The null hypothesis $\beta = 0$ can be tested by a t-test on the differences $\Delta_k = \Theta_{k}^{(i)} - \Theta_{k-1}^{(i)}$. If a trend is detected, $a^{(i)}$ is increased by a fixed factor $f \in \mathbb{R}^+$. 
	
	\item \textbf{Range test (too large $a$):} For each dimension $i = 1, \ldots, p$, $a^{(i)}$ is set to $a^{(i)}/f$ if the trajectory of $\Theta_k^{(i)}$ spans more than $70$\% of the parameter range.

	\item \textbf{Convergence test:} Simulate $n_2$ likelihood estimates at $\Theta_{k-K_0}$ and at $\Theta_{k}$. Growth of the likelihood is then tested by a one-sided Welch's t-test. (A standard hypothesis test; testing for equivalence could be used instead, see \citet{Wellek}.)
\end{itemize}

\end{noname}

We conclude that the algorithm has converged to a maximum only if the convergence test did not reject the null hypothesis three times in a row and at the same time no adjustments of $a$ were necessary. In the applications of the algorithm presented in the article, $f=1.5$.

\subsection{Kernel density estimation}

To enable the estimation of the gradient even far away from the maximum, a kernel function with infinite support is helpful. The Gaussian kernel is an obvious option, but the high rate of decay can by chance cause very large steps leading away from the maximum. Therefore, we use the following modification of the Gaussian kernel: 

\begin{equation*}
	\kappa(H^{1/2}x) \propto \left\{ 
		\begin{array}{ll}
			\exp \left( -\frac{1}{2} x'H^{-1}x \right) & \text{ if }x'H^{-1}x < 1 \\
			\exp \left( - \frac{1}{2} \sqrt{x'H^{-1}x} \right) & \text{ otherwise.}
		\end{array}
	\right.
\end{equation*}

In degenerate cases where the likelihood evaluates to zero numerically, we replace the classical kernel density estimate by a nearest neighbor estimate in step 1d:

\begin{noname}
	If $\hat{L}(\Theta^-; S_{obs}) \approx 0$ or/and  $\hat{L}(\Theta^+; S_{obs}) \approx 0$ (with ``$\approx$'' denoting ``numerically equivalent to''), find 
		\begin{align*}
			S_{\min}^- := \arg \min_{S_j^-}\left\{\left\|S_j^- - S_{obs} \right\|: j=1, \ldots, n\right\} \\
			S_{\min}^+ := \arg \min_{S_j^+}\left\{\left\|S_j^+ - S_{obs} \right\|: j=1, \ldots, n\right\}
		\end{align*}
	and recompute the kernel density estimate $\hat{L}(\Theta^-; s)$ and $\hat{L}(\Theta^+; s)$  in step 1d using $S_{\min}^-$ and  $S_{\min}^+$, respectively, as the only observation.
\end{noname}

To be computationally efficient, we estimate a diagonal bandwidth matrix $H_k$
using a multivariate extension of Silverman's rule \citep{Silverman, HaerdleEtAl}. Using new estimates in each iteration introduces an additional level of noise that can be reduced by using a moving average of bandwidth estimates.

\subsection{Starting points}

To avoid starting in very unfavourable regions of the likelihood, a set of random points should be drawn from the parameter space and their simulated likelihood values compared to find useful starting points. This strategy also helps to avoid that the algorithm reaches only a local maximum.

\subsection{Constraints}

The parameter space will usually be subject to constraints (e.g., rates are positive quantities). They can be incorporated by projecting the iterate to the closest point such that both $\Theta^-$ as well as $\Theta^+$ are in the feasible set \citep{Sadegh97}. Even if there are no imperative constraints, it is advisable to restrict the parameter space to a range of plausible values to prevent the algorithm from trailing off at random in regions of very low likelihood. 

To reduce the effect of large steps within the boundaries, we clamp the step size at 10\% of the range of the feasible set in each dimension.

\section{Examples}
\label{sec:examples}

To study the performance of the AML algorithm, we test it on different applications. The first example is the multivariate normal distribution. While there is no need for simulation based inference for normal models, it allows us to compare the properties of the AML estimator and the maximum likelihood estimator. Second, we apply the algorithm to a queuing process where the exact evaluation of the likelihood can require a prohibitive number of steps. The third example is from population genetics. Here, we use both simulated data, where the true parameter values are known, as well as DNA sequence data from a sample of Bornean and Sumatran orang-utans and estimate parameters of their ancestral history.

\subsection{Multivariate normal distribution}

When the AML algorithm is used to estimate the mean vector $\Theta=(\mu_1, \ldots, \mu_{10})$ of a 10-dimensional normal distribution with a diagonal VC matrix using the arithmetic means $\bar{X} = (\bar{X}_1, \ldots, \bar{X}_{10})$ as summary statistics, convergence is achieved quickly and the results are extremely accurate. 

One dataset is simulated under a 10-dimensional normal distribution such that the maximum likelihood estimator for $\Theta$ is $\hat{\Theta}_{\text{ML}} = \bar{X} \sim \mathcal{N}(5 \cdot \mathbf{1}_{10}, I_{10})$ where $I_{10}$ denotes the $10$-dimensional identity matrix and $\mathbf{1}_{10}$ the $10$-dimensional vector with 1 in each component. To estimate the distribution of $\hat{\Theta}_{\text{AML}}$, the AML algorithm is run 1000 times on this dataset with summary statistics $S = \bar{X}$. 

\begin{figure*}[htbp]
\centering
\includegraphics[width=1\textwidth]{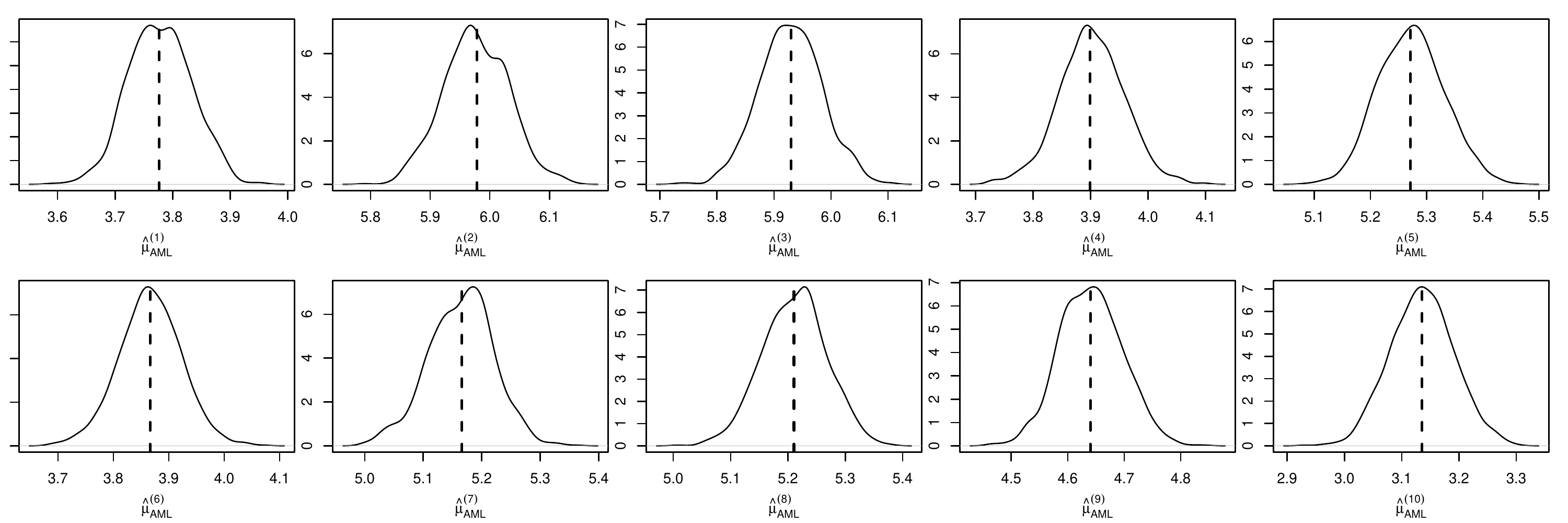}
\caption{Density of the components of $\hat{\Theta}_{\text{AML}}$ obtained with $S=\bar{X}$ in one dataset estimated from 1000 converged sequences with a miminum length of 10000 iterations by kernel density estimation. Vertical dashed line: $\hat{\Theta}_{\text{ML}}$.}
\label{fig:normal-lin}
\end{figure*}

For each AML estimate, 1000 points are drawn randomly on $(-100, 100) \times \ldots \times (-100, 100)$. For each of them, the likelihood is simulated and the 5 points with the highest likelihood estimate are used as starting points. On each of them, the AML algorithm is run for at least 10000 iterations and stopped as soon as convergence is reached (for $\approx 90$\% of the sequences within 11000 iterations; convergence is tested every $K_0=1000$ iterations). Again, the likelihood is simulated on each result and the one with the highest likelihood is considered as a realization of $\hat{\Theta}_{\text{AML}}$. For each evaluation of the likelihood, $n=100$ datasets are simulated. Based on these 1000 realizations of $\hat{\Theta}_{\text{AML}}$, the density, bias and standard error of $\hat{\Theta}_{\text{AML}}$ are estimated for each dimension (tab.~\ref{tab:normal}, fig.~\ref{fig:normal-lin}). 

The densities of the 10 components of $\hat{\Theta}_{\text{AML}}$ are symmetric around the maximum likelihood estimate with a standard error that is nearly 20 times smaller than the standard error of $\hat{\Theta}_{\text{ML}}$ and a negligible bias. Bootstrap confidence intervals that were obtained from $B=100$ bootstrap samples for 100 datasets simulated under the same model as above show a close approximation to the intended coverage probability of $95$\%.

To investigate the impact of the choice of summary statistics on the results, we repeat the experiment with the following set of summary statistics:

\begin{equation}
\label{eq:transformed}
	\begin{array}{ccccc}
	\bar{X}_1 & \bar{X}_2 + \bar{X}_3 & \bar{X}_4 + \bar{X}_5 & \bar{X}_7 & \bar{X}_9 \\
	& \bar{X}_2 - \bar{X}_3 & \bar{X}_5 + \bar{X}_6 & \bar{X}_7 + \bar{X}_8 &\bar{X}_9 \cdot \bar{X}_{10} \\
	& & \bar{X}_6 + \bar{X}_4 & &
	\end{array}
\end{equation}

\begin{figure*}[htbp]
\centering
\includegraphics[width=1\textwidth]{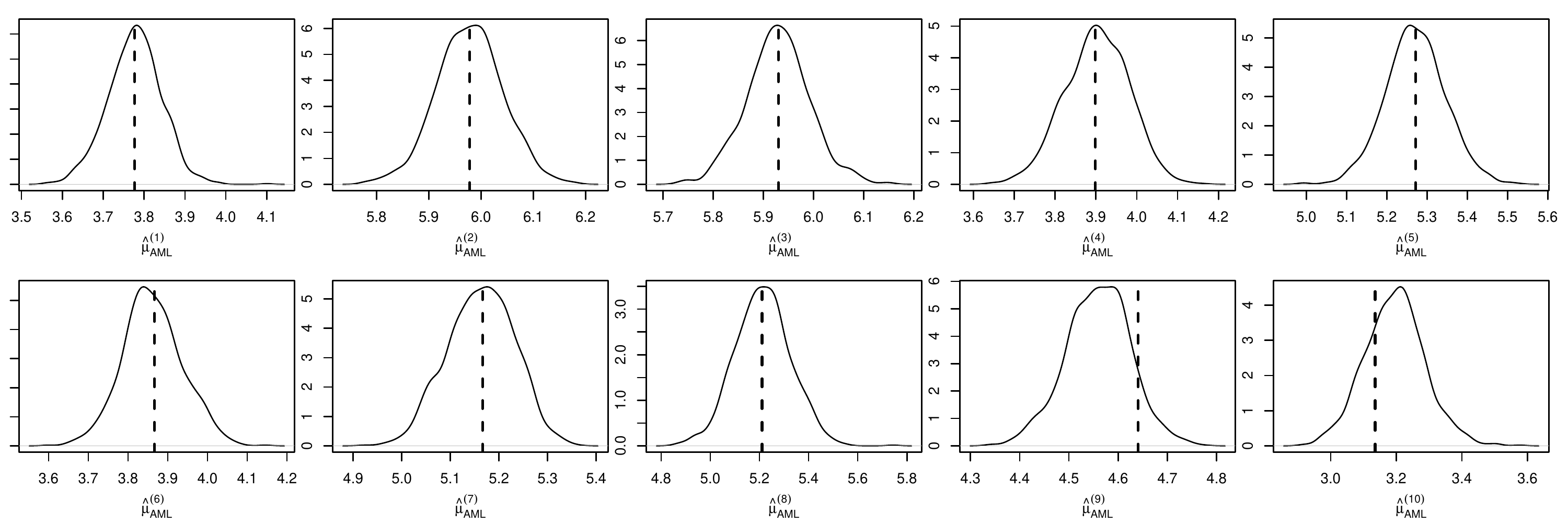}
\caption{Density of the components of $\hat{\Theta}_{\text{AML}}$ obtained with $S$ as in eq. \eqref{eq:transformed} in one dataset estimated from 1000 converged sequences with a miminum length of 10000 iterations by kernel density estimation. Vertical dashed line: $\hat{\Theta}_{\text{ML}}$.}
\label{fig:normal-nl}
\end{figure*}

\begin{figure}[htbp]
\centering
\includegraphics[width=0.5\textwidth]{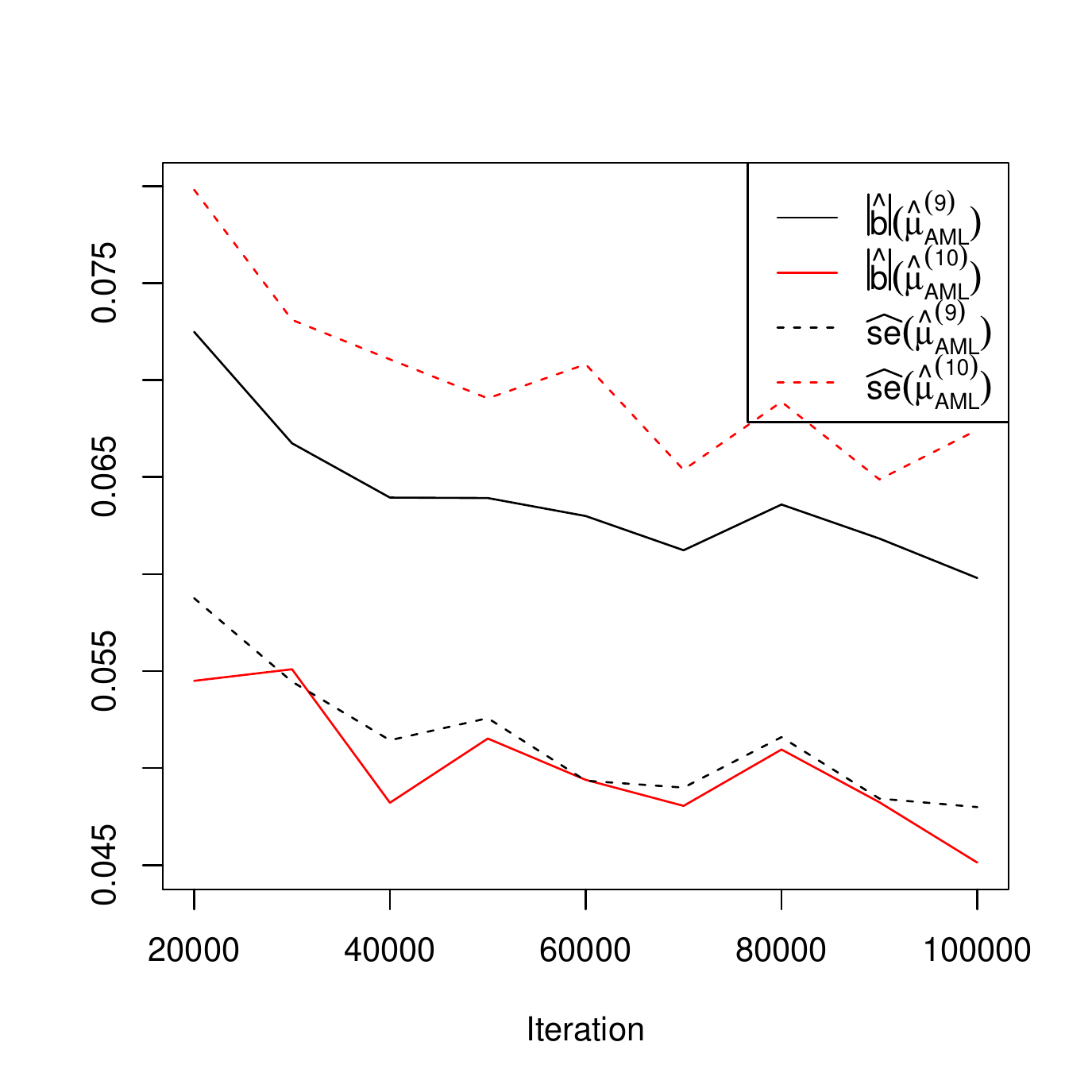}
\caption{Absolute bias and standard error of the AML estimator of $\mu_9$ and $\mu_{10}$ using the summary statistics in eq.~\eqref{eq:transformed} estimated from 1000 runs of the algorithm on the same dataset.}
\label{fig:normal-bias}
\end{figure}

For $\approx 90$\% of the sequences, convergence was detected within 14000 iterations. Bootstrap confidence intervals are obtained for 100 datasets. Their coverage probability matches the nominal 95\% confidence level closely (tab.~\ref{tab:normal}). The components behave very similarly to the previous simulations, except for the estimates of the density for components 9 and 10 (fig.~\ref{fig:normal-nl}). Compared to the simulations with $S=\bar{X}$, the bias of components 9 and 10 is considerably increased, but it is still much smaller than the standard error of $\hat{\Theta}_{\text{ML}}$. To investigate how fast the bias decreases with the number of iterations, we re-run the above described algorithm for 100000 iterations without earlier stopping (fig.~\ref{fig:normal-bias}). Both bias and standard error decrease with the number of iterations.

\begin{table}[ht]
\begin{center}
\caption{Properties of $\hat{\Theta}_{\text{AML}}$ in the 10-dimensional normal distribution model using 2 different sets of summary statistics}
\label{tab:normal}
\begin{tabular}{r|rrr|rrr}
	\hline \noalign{\smallskip}
	& \multicolumn{3}{c|}{$S = \bar{X}$} & \multicolumn{3}{c}{$S$ as in eq. \ref{eq:transformed}} \\
  \hline \noalign{\smallskip}
dim. & $\hat{b}$ & $\widehat{se}$ & $\hat{p}$ & $\hat{b}$ & $\widehat{se}$ & $\hat{p}$ \\ 
  \hline
1 & -0.0016 & 0.0546 & 93 & -0.0007 & 0.0660 & 93 \\ 
  2 & 0.0017 & 0.0544 & 94 & -0.0002 & 0.0638 & 93 \\ 
  3 & 0.0004 & 0.0547 & 94 & 0.0007 & 0.0635 & 97 \\ 
  4 & -0.0033 & 0.0567 & 90 & -0.0032 & 0.0789 & 98 \\ 
  5 & -0.0015 & 0.0584 & 95 & -0.0000 & 0.0748 & 90 \\ 
  6 & -0.0000 & 0.0565 & 94 & 0.0044 & 0.0757 & 90 \\ 
  7 & 0.0017 & 0.0557 & 96 & 0.0035 & 0.0686 & 95 \\ 
  8 & -0.0007 & 0.0559 & 95 & -0.0030 & 0.1140 & 96 \\ 
  9 & -0.0013 & 0.0554 & 91 & 0.0809 & 0.0658 & 98 \\ 
  10 & 0.0001 & 0.0554 & 99 & -0.0595 & 0.0922 & 92 \\ 
   \hline
\end{tabular}
\end{center}
Bias ($\hat{b}$) and standard error ($\widehat{se}$) of $\hat{\Theta}_{\text{AML}}$, estimated from 1000 runs of the AML algorithm. Coverage probability of bootstrap 95\% confidence intervals ($\hat{p}$) with $B=100$, estimated from 100 datasets. 
\end{table}


\subsection{Queueing process} 

The theory of queueing processes is a very prominent and well-studied field of applied probability with applications in many different areas like production management, communication and information technology, and health care \citep{Kingman09, GrossEtAl}. Here, we consider a first-come-first-serve queuing process with a single server, a Markovian distribution of interarrival times of customers and a general distribution of service times (M/G/1). It is modelled by the following sequences of random variables: The service time $U_m$, the interarrival time $W_m$ and the interdeparture time $Y_m$ for customer $m=1, \ldots, M$ \citep{HegglandFrigessi04}. $U_m$ and $W_m$ are independent and their distributions are known except for the parameters. Here we assume $U_m \sim U\left(\left[\theta_1, \theta_1 + \theta_2\right]\right)$ and $W_m \sim \text{exp}(\theta_3)$. $Y_m$ is defined as
\begin{equation*}
	Y_m = \left\{ \begin{array}{ll}
		U_m & \text{if } \sum\limits_{i=1}^{m} W_i \leq \sum\limits_{i=1}^{m-1} Y_i, \\
		U_m + \sum\limits_{i=1}^{m} W_i - \sum\limits_{i=1}^{m-1} Y_i & \text{otherwise}
	\end{array} \right.
\end{equation*}
for $m=1, \ldots, M$. 

If only the interdeparture times $Y_m$ are observed, the evaluation of the likelihood of the parameter vector $\Theta = (\theta_1, \theta_2, \theta_3)$ requires an exponential number of steps in $M$ \citep{HegglandFrigessi04}. Approximate inference has been conducted with indirect inference \citep{HegglandFrigessi04} and ABC methods \citep{BlumFrancois10}. 

To estimate the distribution of $\hat{\Theta}_{\text{AML}}$ in this model, we simulate 100 datasets of size $M=100$ under the same parameter value $\Theta = (1, 4, 0.2)$ and compute $\hat{\Theta}_{\text{AML}}$ for each of them. 
The summary statistics are the minimum, maximum and the quartiles of $Y_1, \ldots, Y_M$ \citep{BlumFrancois10}.
To estimate the standard error of $\hat{\Theta}_{\text{AML}}$ per dataset, we re-run the algorithm five times on each dataset. 

Each time, we randomly draw 100 starting values from the search space $(0,10) \times (0,10) \times (0.05, 10)$ and run the AML algorithm on the best five starting values for $5000$ iterations (runtime of an implementation in R, version 3.0.1 \cite{R2013}: $1.9$ hours on a single core). Finally, we estimate the likelihood on the 5 results and the best one is used as a realization of $\hat{\Theta}_{\text{AML}}$. Each likelihood estimation is based on $n=100$ simulations. For 3 datasets, the AML algorithm is run 100 times to obtain kernel density estimates of the density of $\hat{\Theta}_{\text{AML}}$. 

\setlength{\tabcolsep}{5pt}

\begin{table}[htbp]
\begin{center}
\caption{Properties of $\hat{\Theta}_{\textrm{AML}}$ in the M/G/1 queue}
\label{tab:gg1}
\begin{tabular}{llll}
   \hline
\multirow{7}{*}{$\mathrm{\boldsymbol{\theta_1}}$} & \multirow{7}{*}{\includegraphics[width=4cm]{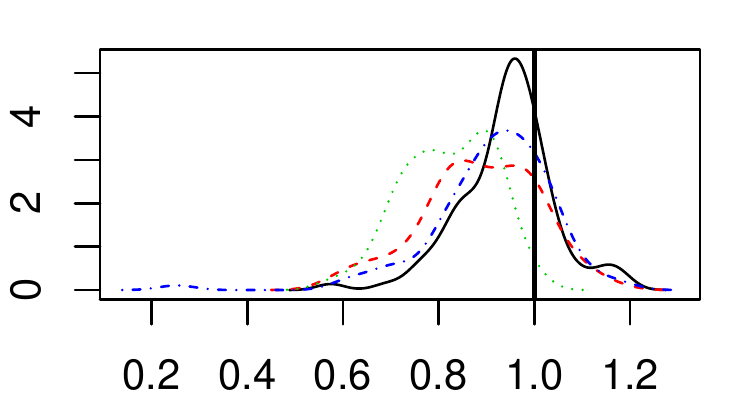}} & true & 1 \\ 
   &  & space & $(0, 10)$ \\ 
   &  & mean & 0.942 \\ 
   &  & median & 0.948 \\ 
   &  & bias & -0.058 \\ 
   &  & mean se & 0.062 \\ 
   &  & total se & 0.098 \\ 
   \hline
\multirow{7}{*}{$\mathrm{\boldsymbol{\theta_2}}$} & \multirow{7}{*}{\includegraphics[width=4cm]{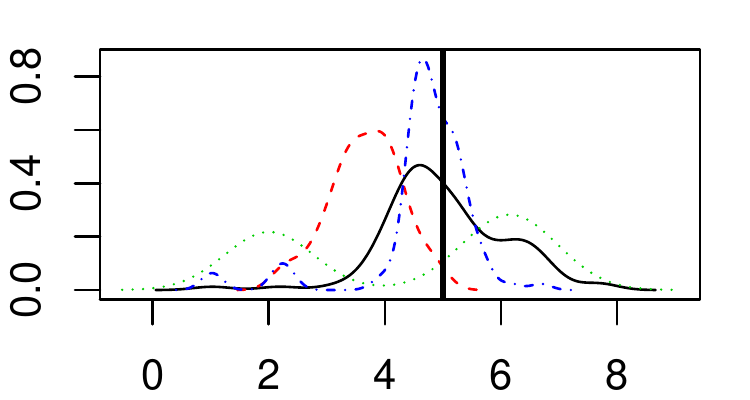}} & true & 5 \\ 
   &  & space & $(0, 10)$ \\ 
   &  & mean & 5.031 \\ 
   &  & median & 4.891 \\ 
   &  & bias & 0.031 \\ 
   &  & mean se & 0.437 \\ 
   &  & total se & 1.046 \\ 
   \hline
\multirow{7}{*}{$\mathrm{\boldsymbol{\theta_3}}$} & \multirow{7}{*}{\includegraphics[width=4cm]{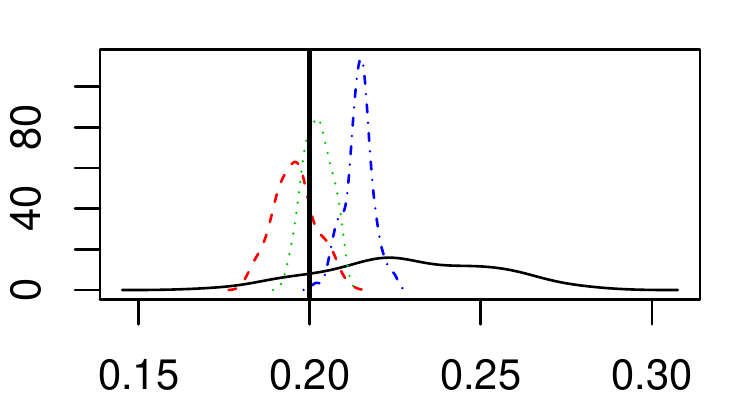}} & true & 0.2 \\ 
   &  & space & $(0.05, 10)$ \\ 
   &  & mean & 0.229 \\ 
   &  & median & 0.226 \\ 
   &  & bias & 0.029 \\ 
   &  & mean se & 0.004 \\ 
   &  & total se & 0.024 \\ 
   \hline
\end{tabular}
\end{center}

Figures: marginal densities of components of $\hat{\Theta}_{\textrm{AML}}$, estimated by kernel density estimation. Black solid line: density of $\hat{\Theta}_{\textrm{AML}}$ estimated over 100 simulated data sets. Coloured dotted and dashed lines: density of $\hat{\Theta}_{\textrm{AML}}$ estimated separately in 3 example datasets. Vertical black line: true parameter value. 

Summaries: {\it true:} true parameter value; {\it space:} search space; {\it mean:} mean of $\hat{\Theta}_{\textrm{AML}}$; {\it median:} median of $\hat{\Theta}_{\textrm{AML}}$; {\it bias:} bias of $\hat{\Theta}_{\textrm{AML}}$; {\it mean se:} mean standard error of $\hat{\Theta}_{\textrm{AML}}$ per dataset; {\it total se:} standard error of $\hat{\Theta}_{\textrm{AML}}$.
\end{table}
\setlength{\tabcolsep}{6pt}

In all three dimensions, the standard error of $\hat{\Theta}_{\text{AML}}$ is considerably larger across datasets than within dataset (tab.~\ref{tab:gg1}). This indicates that the approximation algorithm only adds a small additional error to the error of $\hat{\Theta}_{\text{ML}}$. In all dimensions the density of $\hat{\Theta}_{\text{AML}}$ is slightly asymmetric around the true value, but only the bias of $\hat{\theta}_3$ is of the same order as the standard error. It is probably caused by a strongly skewed likelihood function \citep{BlumFrancois10} that emphasises the error of the finite differences approximation to the slope. Additional simulations of the likelihood on a grid between $\hat{\theta}_3$ and $\theta_3$ locate the maximum around $0.209$. Re-running the algorithm for $K=10000$ iterations reduces the bias from $0.029$ to $0.024$. The parameter estimate with the largest standard error is $\hat{\theta}_2$ which is also the most difficult parameter to estimate with ABC \citep{BlumFrancois10}.

\subsection{Population genetics}

The goal of population genetics is to understand how evolutionary forces like mutation, recombination, selection and genetic drift shape the variation within a population. A major challenge in the application of population genetic theory is to infer parameters of the evolutionary history of a population from DNA sequences of present-day samples only. The questions range from estimating single parameters like the mutation rate or the recombination rate to determining a complex demographic structure with multiple parameters. 

Since the introduction of the coalescent by Kingman in the early eighties \citep{Kingman82a, Kingman82b, Kingman82c}, a very elegant and flexible stochastic process is at hand to model the ancestral history of a population. The coalescent can capture population structure, varying population size and different mating schemes. Extensions to incorporate recombination and selection exist as well \citep{Nordborg07, Wakeley}. 

However, statistical inference on data obtained from a coalescent process is difficult: Under the coalescent, computing the likelihood of a parameter vector is a computationally very expensive task even for a single locus, since all possible ancestral trees need to be taken into account \citep{Stephens07}. The number of binary tree topologies explodes with the number of sampled individuals, $n$, e. g. with $n=10$, the number of possible topologies is $2,571,912,000$, with $n=1000$, it is already $3.01 \cdot 10^{4831}$ \citep[p. 83, table 3.2]{Wakeley}.

Consequently, exact computation of the likelihood is feasible in reasonable time for a very restricted set of models and a small number of haplotypes only \citep{GriffithsTavare94a, Wu10}. To make use of the large DNA data sets generated by modern sequencing technology, approximate methods have been developed in frequentist as well as in Bayesian frameworks. In fact, the need  for inferential tools that allow for large genomic data sets was a major driver in the development of ABC. The focus of the early ABC papers was clearly on population genetic applications \citep{WeissVonHaeseler98, Pritchard99, BeaumontEtAl02} and since then it has found a large variety of applications in the field of population genetics \citep{CsilleryEtAl10, Beaumont10, BertorelleEtAl10}.

\subsubsection{Population genetic history of orang-utans}

\textit{Pongo pygmaeus} and \textit{Pongo abelii}, Bornean and Sumatran orang-utans, respectively, are Asian great apes whose distributions are exclusive to the islands of Borneo and Sumatra. Recurring glacial periods that led to a cooler, drier and more seasonal climate might have contracted rain forest and isolated the population of orang-utans. At the same time, the sea level dropped and land bridges among islands created opportunities for migration among previously isolated populations. However, whether glacial periods have been an isolating or a connecting factor remains poorly understood. Therefore, there has been a considerable interest in using genetic data to understand the demographic history despite the computational difficulties involved in such a population genetic analysis. We will compare our results  to the analysis of the orang-utan genome paper \citep{LockeEtAl11} and a more comprehensive study by Ma et al.~\citep{MaXinEtAl13}; the only analyses that have been performed genome-wide. Both studies use DaDi~\citep{GutenkunstEtAl09}, a state-of-art software that has been widely used for demographic inference. It is based on the diffusion approximation to the coalescent. The orang-utan SNP data consisting of 4-fold degenerate (synonymous) sites are taken from the 10 sequenced individuals (5 individuals each  per orang-utan population, haploid sample size 10, ~\citealp{LockeEtAl11}).

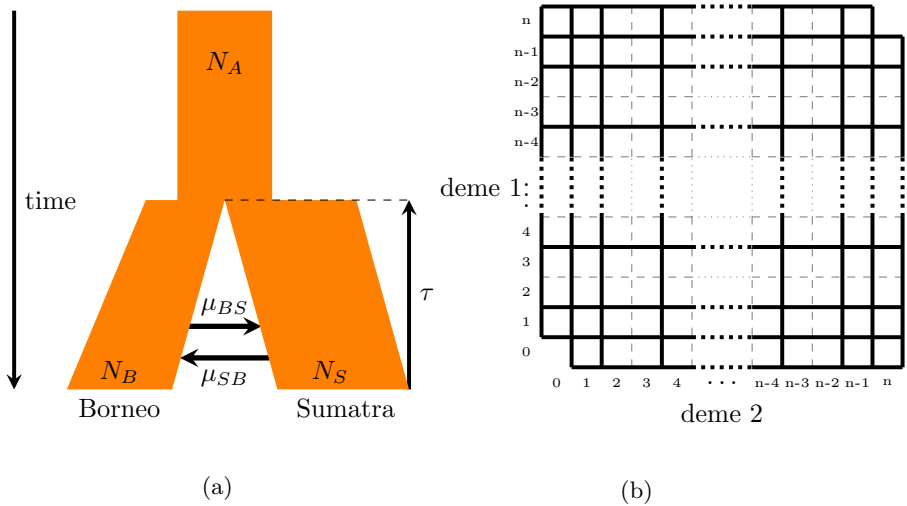
\begin{figure*}[htbp]%
\centering
\begin{subfigure}{0.45\linewidth}
	\centering
	\begin{tikzpicture}[scale=0.7, yscale=1.2]
		\draw[->,line width=2pt,>=stealth] (1.8,1) -- (3.2,1) node [pos=0.5,above] {$\mu_{B S}$};
		\draw[<-,line width=2pt,>=stealth] (1.65,.5) -- (3.35,.5) node [pos=0.5,below] {$\mu_{S B}$};
		\draw[->,line width=1.5pt,>=stealth] (6,0) -- (6,3) node [pos=0.5,right] {$\tau$};
		\shade[top color=orange,bottom color=orange,thick]
			(-0.5,0) -- (1.5,0) -- (2.5,3) -- (3.5,0) -- (6,0) --
			(5,3) -- (3.4, 3) -- (3.4,6) -- (1.6,6) -- (1.6,3) --(1,3)-- (-0.5,0)
			(0.5,0.6) node[below] {$N_{B}$}
			(0.5,0) node[below] {Borneo}
			(4.5,0.6) node[below] {$N_{S}$}
			(4.8,0) node[below] {Sumatra}
			(2.5,5.5) node[below] {$N_A$};
		\draw[dashed] (2.5,3) -- (6,3);
		\draw[<-, line width=1.5pt,>=stealth] (-1.5, 0) -- (-1.5, 6) node [pos=0.5,right] {time};
	\end{tikzpicture}
	\caption{}
	\label{fig:IM}
\end{subfigure}
\begin{subfigure}{0.45\linewidth}
	\centering
	\begin{tikzpicture}
[scale=0.4, 
basic/.style={dashed, gray},
binned/.style={line width=1.5pt}]
	
	\foreach \x in {3,5,7,9}
	{
		\draw[basic] (\x,0) -- (\x,5);
		\draw[basic, dotted] (\x,5) -- (\x,7);
		\draw[basic] (\x,7) -- (\x,12);
		\draw[basic] (0,\x) -- (5,\x);
		\draw[basic, dotted] (5,\x) -- (7,\x);
		\draw[basic] (7,\x) -- (12,\x);
	}
	
	\foreach \x in {1,2,4,8,10,11}
	{
		\draw[binned] (\x,0) -- (\x,5);
		\draw[binned, dotted] (\x,5) -- (\x,7);
		\draw[binned] (\x,7) -- (\x,12);
		\draw[binned] (0,\x) -- (5,\x);
		\draw[binned, dotted] (5,\x) -- (7,\x);
		\draw[binned] (7,\x) -- (12,\x);
	}
	\draw[binned] (0,1) -- (0,5);
	\draw[binned, dotted] (0,5) -- (0,7);
	\draw[binned] (0,7) -- (0,12);
	
	\draw[binned] (1,0) -- (5,0);
	\draw[binned, dotted] (5,0) -- (7,0);
	\draw[binned] (7,0) -- (12,0);
	
	\draw[binned] (12,0) -- (12,5);
	\draw[binned, dotted] (12,5) -- (12,7);
	\draw[binned] (12,7) -- (12,11);
	
	\draw[binned] (0,12) -- (5,12);
	\draw[binned, dotted] (5,12) -- (7,12);
	\draw[binned] (7,12) -- (11,12);
	
	\foreach \x in {0, ..., 4}{
		\draw (\x+0.5, -0.5) node{\tiny{\x}};
		\draw (-0.5, \x+0.5) node{\tiny{\x}};
	}
	\foreach \x in {1, ..., 4}{
		\draw (12-\x-0.5, -0.5) node{\tiny{n-\x}};
		\draw (-0.5, 12-\x-0.5) node{\tiny{n-\x}};
	} 
	\draw (12-0.5, -0.5) node{\tiny{n}};
	\draw (-0.5, 12-0.5) node{\tiny{n}};
	\draw (6, -0.5) node{$\ldots$};
	\draw (-0.5, 6) node{$\vdots$};
	
	\draw (6, -1.5) node{deme 2};
	\draw (-2, 6) node{deme 1};	
	
\end{tikzpicture}
	\caption{}
	\label{fig:JSFS}
\end{subfigure}
\caption{(a) Isolation-migration model for the ancestral history of orang-utans. $N_A$, effective size of the ancestral population; $\mu_{BS}$ ($\mu_{SB}$), fraction of the Bornean (Sumatran) population that is replaced by Sumatran (Bornean) migrants per generation (backwards migration rate); $\tau$, split time in years; $N_B$ ($N_S$), effective population size in Borneo (Sumatra).\\ (b) Binned joint site frequency spectrum \citep[adapted from][]{NaduvilezhathEtAl11}.}
\end{figure*}

As in \cite{LockeEtAl11} and \cite{MaXinEtAl13}, we consider an Isolation-Migration (IM) model where a panmictic ancestral population of effective size $N_A$ splits $\tau$ years ago into two distinct populations of constant effective size $N_B$ (the Bornean population) and $N_S$ (the Sumatran population) with backward migration rates $\mu_{BS}$ (fraction of the Bornean population that is replaced by Sumatran migrants per generation) and $\mu_{SB}$ (vice versa; fig.~\ref{fig:IM}). 

$N_A$ is set to the present effective population size that we obtain using the number of SNPs in our considered data set and assuming an average per generation mutation rate per nucleotide of $2 \cdot 10^{-8}$ and a generation time of 20 years \citep{LockeEtAl11}, so $N_A = 17400$. 

There are no sufficient summary statistics at hand, but for the IM model the joint site frequency spectrum (JSFS) between the two populations was reported to be a particularly informative summary statistic \citep{TellierEtAl11}. However, for $N$ samples in each of the two demes, the JSFS has $(N+1)^2-2$ entries, so even for small datasets it is very high-dimensional. To reduce this to more manageable levels we follow \citet{NaduvilezhathEtAl11} and bin categories of entries (fig.~\ref{fig:JSFS}). As the ancestral state is unknown, we use the folded binned JSFS that has 28 entries. 
To incorporate observations of multiple unlinked loci mean and standard deviation across loci are computed for each bin, so the final summary statistics vector is of length 56. 

The AML algorithm with the JSFS as summary statistics was implemented as an extra module in the \texttt{msms} package, a coalescent simulation program \citep{EwingHermisson10}. This allows for complex demographic models and permits fast summary statistic evaluation without using external programs and the associated performance penalties.

\subsubsection{Simulations}

Before applying the AML algorithm to the actual orang-utan DNA sequences, we tested it on simulated data. Under the described IM model with parameters $N_B=N_S=17400$, $\mu_{BS}= \mu_{SB}=1.44 \cdot 10^{-5}$ and $\tau=695000$, we simulated 25 datasets with 25 haploid sequences per deme, each of them consisting of 75 loci with 130 SNPs each. 

For each dataset, 25 AML estimates were obtained with the same scheme: 1000 random starting points were drawn from the parameter space; the likelihood was estimated with $n=40$ simulations. Then, the 5 values with the highest likelihood estimates were used as starting points for the AML algorithm. The algorithm converged after $3000$-$25000$ iterations (average: $\approx 8000$ iterations; average runtime of \texttt{msms}: $11.7$ hours on a single core). 

For the migration rates $\mu_{SB}$ and $\mu_{BS}$, the per dataset variation of the estimates is considerably smaller than the total variation (tab.~\ref{tab:coal-properties}). This suggests that the approximation error of the AML algorithm is small in comparison to the error of $\hat{\Theta}_{\text{ML}}$. For the split time $\tau$ and the population sizes $N_{S}$ and $N_{B}$, the difference is less pronounced, but still apparent. For all parameters, the average bias of the estimates is either smaller or of approximately the same size as the standard error. As the maximum likelihood estimate itself cannot be computed, it is impossible to disentangle the bias of $\hat{\Theta}_{\text{ML}}$ and an additional bias introduced by the AML algorithm. As an alternative measure of performance, we compare the likelihood estimated at the true parameter values and the AML estimate with the highest estimated likelihood on each dataset. Only in 2 of the 25 datasets, the likelihood at the true value is higher, but not significantly so, whereas it is significantly lower in 12 of them
(significance was tested with a one-sided Welch test using a Bonferroni-correction to account for multiple testing). This suggests that the AML algorithm usually produces estimates that are closer to the maximum likelihood estimate than the true parameter value is. 

\begin{table*}[htbp]
	\begin{center}
	\caption{Properties of $\hat{\Theta}_{\textrm{AML}}$ in the IM model with short divergence time ($\tau = 695000$ years)}
	\label{tab:coal-properties}
\begin{tabular}{llll}
   \hline
\multirow{7}{*}{$\mu_{BS}$} & \multirow{7}{*}{\includegraphics[width=5cm]{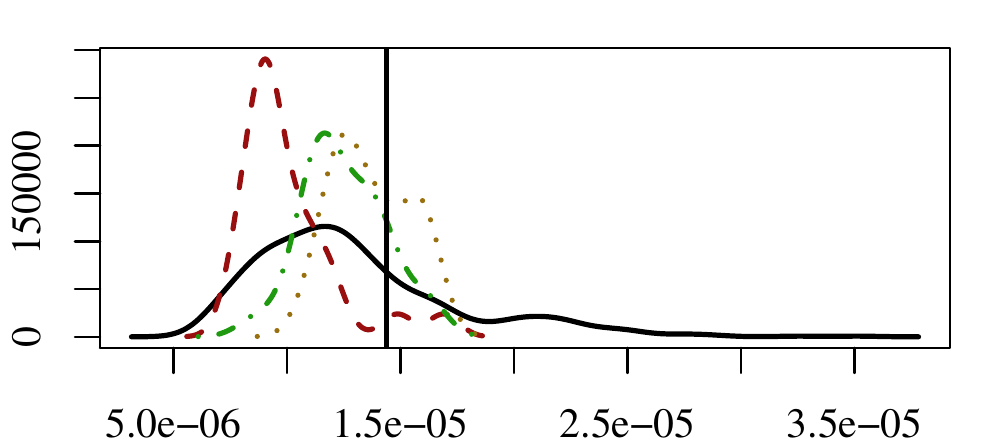}} & true & 1.44e-05 \\ 
   &  & space & (1.44e-06, 0.000144) \\ 
   &  & mean & 1.3e-05 \\ 
   &  & median & 1.21e-05 \\ 
   &  & bias & -1.43e-06 \\ 
   &  & mean se & 2e-06 \\ 
   &  & total se & 4.48e-06 \\ 
   \hline
\multirow{7}{*}{$\mu_{SB}$} & \multirow{7}{*}{\includegraphics[width=5cm]{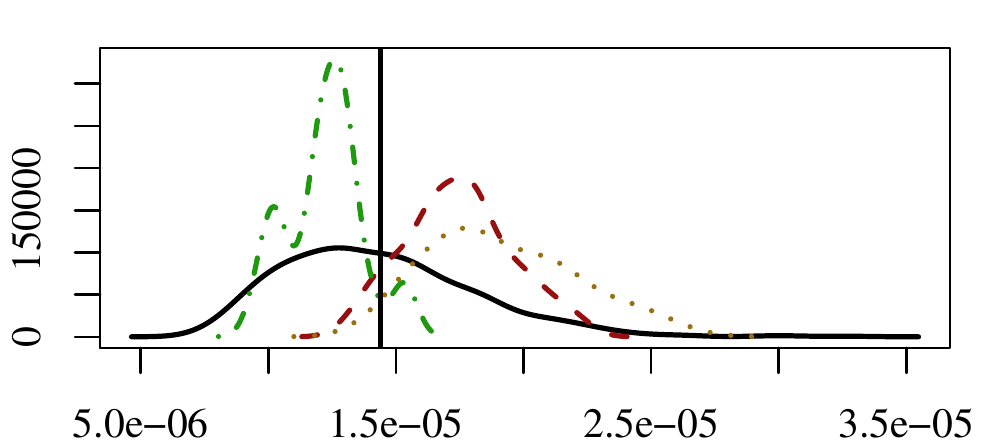}} & true & 1.44e-05 \\ 
   &  & space & (1.44e-06, 0.000144) \\ 
   &  & mean & 1.43e-05 \\ 
   &  & median & 1.39e-05 \\ 
   &  & bias & -7.2e-08 \\ 
   &  & mean se & 2.19e-06 \\ 
   &  & total se & 3.77e-06 \\ 
   \hline
\multirow{7}{*}{$\tau$} & \multirow{7}{*}{\includegraphics[width=5cm]{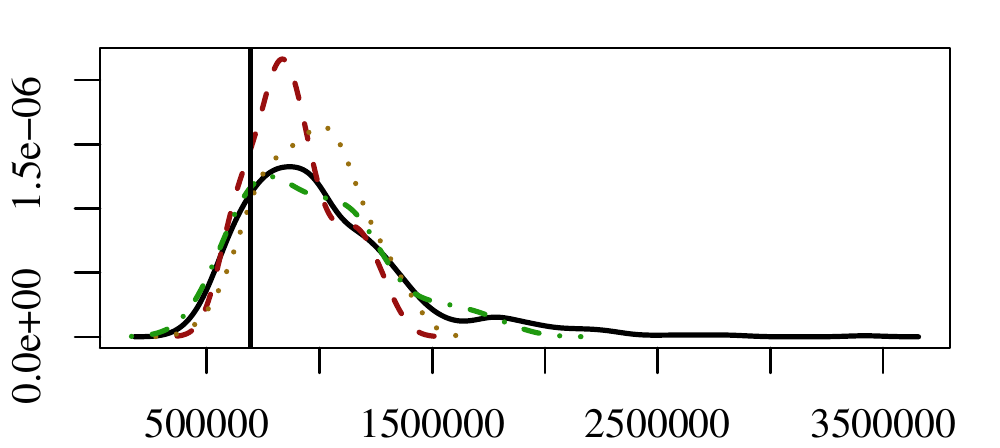}} & true & 695000 \\ 
   &  & space & (139000, 6950000) \\ 
   &  & mean & 1020000 \\ 
   &  & median & 946000 \\ 
   &  & bias & 330000 \\ 
   &  & mean se & 305000 \\ 
   &  & total se & 389000 \\ 
   \hline
\multirow{7}{*}{$N_S$} & \multirow{7}{*}{\includegraphics[width=5cm]{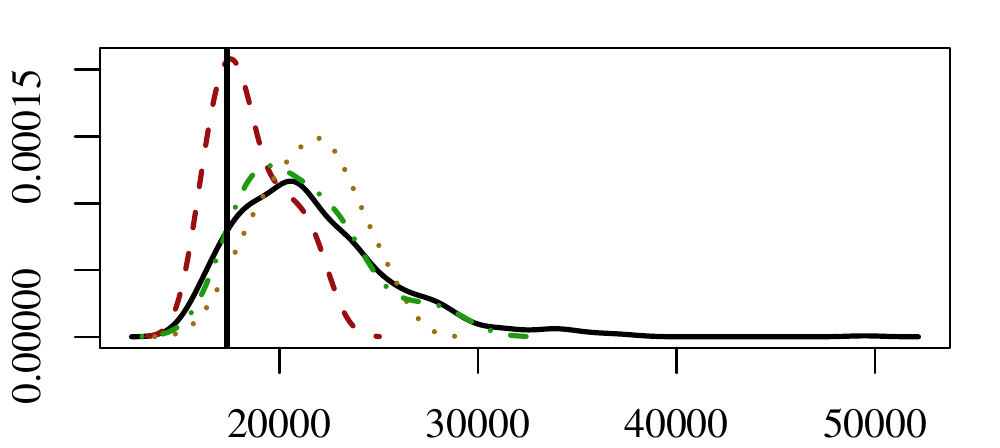}} & true & 17400 \\ 
   &  & space & (1740, 174000) \\ 
   &  & mean & 21700 \\ 
   &  & median & 21000 \\ 
   &  & bias & 4370 \\ 
   &  & mean se & 3070 \\ 
   &  & total se & 4100 \\ 
   \hline
\multirow{7}{*}{$N_B$} & \multirow{7}{*}{\includegraphics[width=5cm]{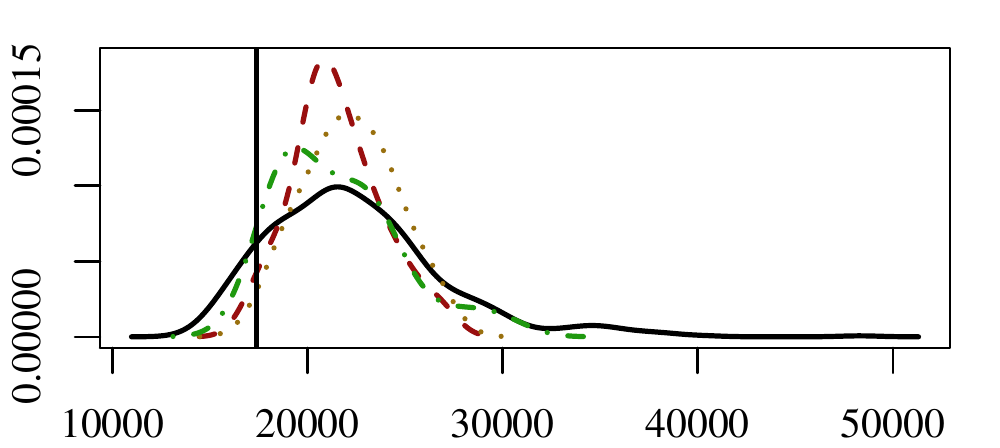}} & true & 17400 \\ 
   &  & space & (1740, 174000) \\ 
   &  & mean & 22400 \\ 
   &  & median & 21800 \\ 
   &  & bias & 5070 \\ 
   &  & mean se & 3160 \\ 
   &  & total se & 4550 \\ 
   \hline
\end{tabular}
	\end{center}
	
	Figures and summaries: As in Tab.~\ref{tab:gg1}.
\end{table*}

%

To investigate the impact of the underlying parameter value on the quality of the estimates, simulation results were obtained also for 25 datasets simulated with $\tau$ twice as large. Here, all parameter estimates, especially $\tau$, $N_{B}$ and $N_{S}$ had larger standard errors and biases (tab.~\ref{tab:coal-properties2}). Apparently, the estimation problem is more difficult for more distant split times. This can be caused by a flatter likelihood surface and by stronger random noise in the simulations. Only the migration rates are hardly affected by the large $\tau$: a longer divergence time allows for more migration events that might facilitate their analysis.

\begin{table*}[htbp]
	\begin{center}
	\caption{Properties of $\hat{\Theta}_{\textrm{AML}}$ in the IM model with long divergence time ($\tau = 1390000$ years)}
	\label{tab:coal-properties2}
\begin{tabular}{llll}
   \hline
\multirow{7}{*}{$\mu_{BS}$} & \multirow{7}{*}{\includegraphics[width=5cm]{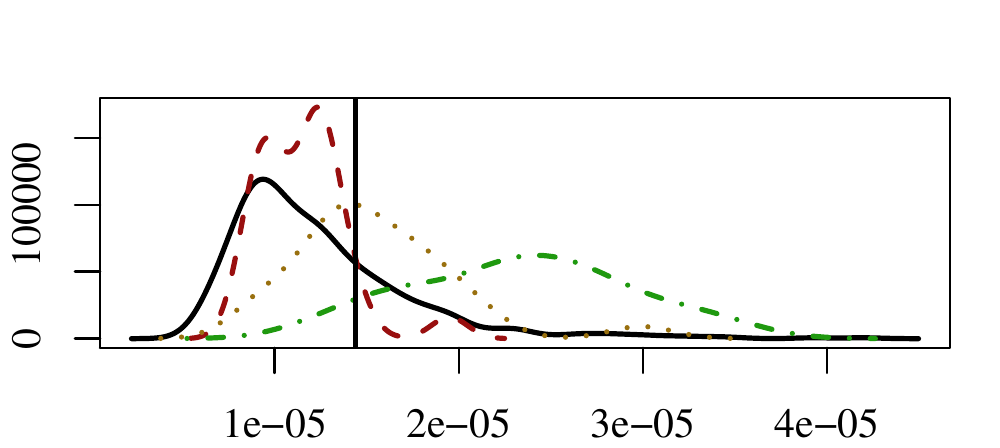}} & true & 1.44e-05 \\ 
   &  & space & (1.44e-06, 0.000144) \\ 
   &  & mean & 1.24e-05 \\ 
   &  & median & 1.12e-05 \\ 
   &  & bias & -1.97e-06 \\ 
   &  & mean se & 3.38e-06 \\ 
   &  & total se & 4.99e-06 \\ 
   \hline
\multirow{7}{*}{$\mu_{SB}$} & \multirow{7}{*}{\includegraphics[width=5cm]{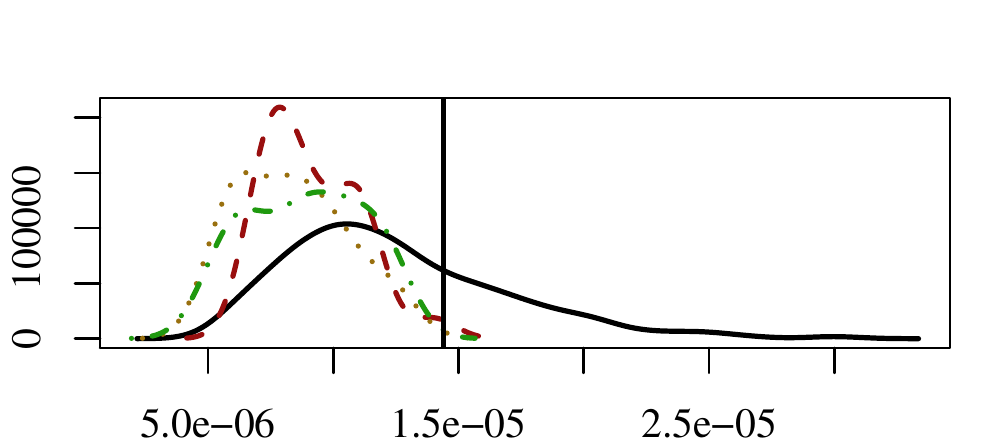}} & true & 1.44e-05 \\ 
   &  & space & (1.44e-06, 0.000144) \\ 
   &  & mean & 1.25e-05 \\ 
   &  & median & 1.17e-05 \\ 
   &  & bias & -1.88e-06 \\ 
   &  & mean se & 3.28e-06 \\ 
   &  & total se & 4.38e-06 \\ 
   \hline
\multirow{7}{*}{$\tau$} & \multirow{7}{*}{\includegraphics[width=5cm]{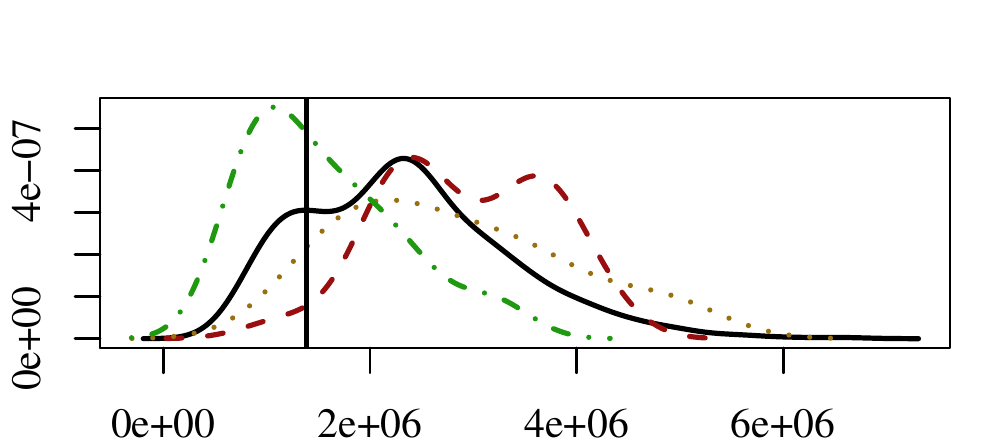}} & true & 1390000 \\ 
   &  & space & (139000, 6950000) \\ 
   &  & mean & 2360000 \\ 
   &  & median & 2280000 \\ 
   &  & bias & 967000 \\ 
   &  & mean se & 876000 \\ 
   &  & total se & 1e+06 \\ 
   \hline
\multirow{7}{*}{$N_S$} & \multirow{7}{*}{\includegraphics[width=5cm]{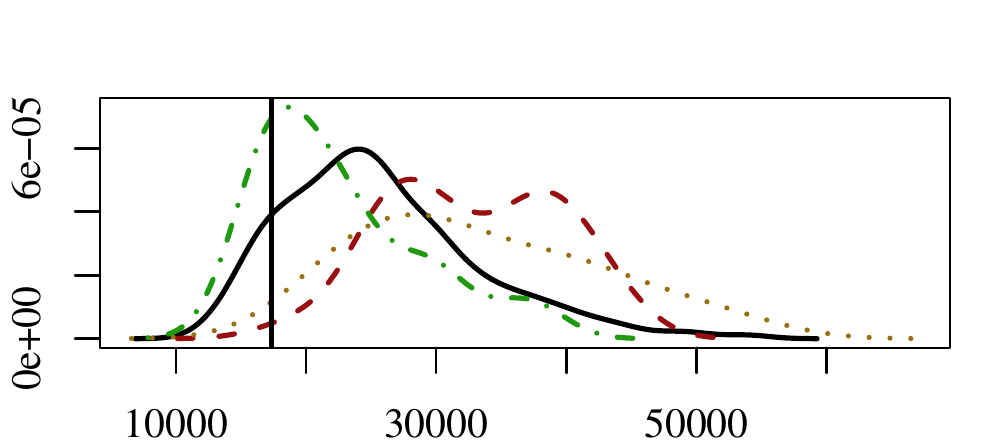}} & true & 17400 \\ 
   &  & space & (1740, 174000) \\ 
   &  & mean & 25700 \\ 
   &  & median & 24600 \\ 
   &  & bias & 8350 \\ 
   &  & mean se & 6620 \\ 
   &  & total se & 7560 \\ 
   \hline
\multirow{7}{*}{$N_B$} & \multirow{7}{*}{\includegraphics[width=5cm]{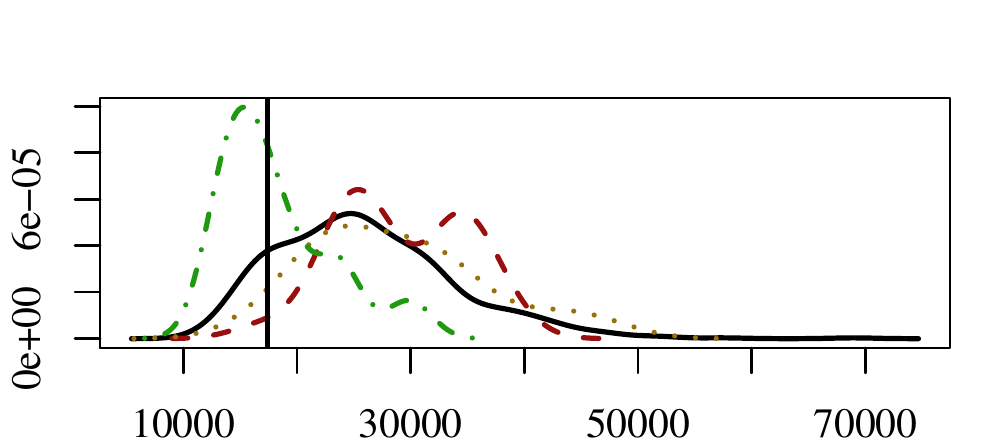}} & true & 17400 \\ 
   &  & space & (1740, 174000) \\ 
   &  & mean & 26000 \\ 
   &  & median & 25200 \\ 
   &  & bias & 8670 \\ 
   &  & mean se & 6680 \\ 
   &  & total se & 7870 \\ 
   \hline
\end{tabular}
	\end{center}
	
	Figures and summaries: As in Tab.~\ref{tab:gg1}.
\end{table*}

%

\subsubsection{Real data}
\label{sec:application}

We model the ancestral history of orang-utans with the same model and use the same summary statistics as tested in the simulations. Also the datasets are simulated in the same manner.

To study the distribution of $\hat{\Theta}_{\text{AML}}$ on this dataset, the algorithm is run 20 times with the same scheme as in the simulations, so we can estimate the standard error of $\hat{\Theta}_{\text{AML}}$ in this single dataset. The best one is used as the final parameter estimate $\hat{\Theta}_{\text{AML}}$ and is used for bootstrapping.

Confidence intervals are obtained by parametric bootstrap with $B=1000$ bootstrap datasets. The bootstrap replicates are also used for bias correction and estimation of the standard error (tab.~\ref{tab:orang}).

\setlength{\tabcolsep}{4pt}
\begin{table}[htbp]
\begin{center}
\caption{Parameter estimates for the ancestral history of orang-utans} 
\label{tab:orang}
\begin{tabular}{rrrrrr}
  \hline
 & $\mu_{BS}$ & $\mu_{SB}$ & $\tau$ & $N_S$ & $N_B$ \\ 
  \hline
$\hat{\Theta}_\textrm{AML}$ & 5.023e-06 & 4.600e-06 & 1300681 & 52998 & 21971 \\ 
  $\hat{\Theta}^\ast_\textrm{AML}$ & 4.277e-06 & 3.806e-06 & 1402715 & 52715 & 22233 \\ 
  $\widehat{se}^\ast$ & 1.244e-06 & 9.366e-07 & 208391 & 7223 & 2779 \\ 
  $\widehat{se}$ & 1.992e-07 & 1.066e-07 & 194868 & 6083 & 2963 \\ 
  lower & 0 \footnotemark[1] & 1.633e-07 & 715590 & 31476 & 13290 \\ 
  upper & 6.627e-06 & 4.931e-06 & 1820852 & 67118 & 27426 \\ 
   \hline
\end{tabular}
 \\ \footnotesize{\footnotemark[1]This confidence interval was shifted to the positive range. The original value was (-9.09e-07,5.72e-06).}
\end{center}
$\hat{\Theta}_{\text{AML}}$, approximate maximum likelihood estimate; $\hat{\Theta}_{\text{AML}}^\ast$, bootstrap bias corrected estimate; $\widehat{se}^\ast$, bootstrap standard error of $\hat{\Theta}_{\text{AML}}$; $\widehat{se}$, standard error of $\hat{\Theta}_{\text{AML}}$ in this dataset, estimated from 20 replicates of $\hat{\Theta}_{\text{AML}}$; lower and upper limits of the $95$\% simultaneous bootstrap confidence intervals. All bootstrap results were obtained with $B=1000$ bootstrap replicates. The simultaneous $95$\% confidence intervals are computed following a simple Bonferroni argument, so the coverage probabilities are $99$\% in each dimension \citep{DavisonHinkley}.
\end{table}

In \citet{LockeEtAl11} and \citet{MaXinEtAl13}, parameters of two different IM models are estimated, denote the estimates $\hat{\Theta}_1$ and $\hat{\Theta}_2$. Scaled to the ancestral population size $N_A = 17372$, the estimates are shown in tab.~\ref{tab:locke}.

\setlength{\tabcolsep}{4pt}
\begin{table}
	\centering
	\caption{Results from \cite[Tab.~S21-1]{LockeEtAl11} (same results for Model 2 reported in \cite{MaXinEtAl13}), scaled with $N_e = 17400$.}
	\label{tab:locke}
\begin{tabular}{rrrrrr}
  \hline
 & $\mu_{BS}$ & $\mu_{SB}$ & $\tau$ & $N_S$ & $N_B$ \\ 
  \hline
Model 1 & 9.085e-07 & 7.853e-07 & 6948778 & 129889 & 50934 \\ 
  Model 2 & 1.518e-05 & 2.269e-05 & 630931 & 35976 & 10093 \\ 
  $\hat{\Theta}_\textrm{AML}$ & 5.023e-06 & 4.600e-06 & 1300681 & 52998 & 21971 \\ 
   \hline
\end{tabular}
	
	Model 1: IM model in figure \ref{fig:IM}. Model 2: IM-model where the ancestral population splits in two subpopulations with a ratio of $s=0.503$ (estimated) going to Borneo and $1-s$ to Sumatra and exponential growth in both subpopulations \cite[Fig.~S21-3]{LockeEtAl11}. Here, $N_B$ and $N_S$ are the present population sizes. 
\end{table}

Model 1 is identical to the model considered here, so we simulate the likelihood at $\hat{\Theta}_1$ within our framework for comparison. Since $\log \hat{L} (\hat{\Theta}_1) = -217.015$ ($se = 7.739$) is significantly lower than $\log \hat{L} (\hat{\Theta}_{\text{AML}}) = -162.732$ ($se = 7.258$), it seems that $\hat{\Theta}_{\text{AML}}$ is closer to the maximum likelihood estimate than the competing estimate. Note, however, that we are only using a subset of the data to avoid sites under selection and that the authors report convergence problems of DaDi in this model. 

For model 2, the ancestral population splits in two subpopulations of size $s$ and $1-s$ relative to the ancestral population and the subpopulations experience exponential growth. Here a direct likelihood comparison with the DaDi estimates given in \cite{LockeEtAl11} is impossible. However, a rough comparison shows that the AML estimates for $\tau$, $N_B$ and $N_S$ lie between $\hat{\Theta}_1$ and $\hat{\Theta}_2$ and for $\mu_{BS}$ and $\mu_{SB}$ they are of similar size.

\subsubsection{Orang-utan data set}

This real data is based on data sets of two publications, \cite{DeMaioEtAl13} and \cite{MaXinEtAl13}. For the first, CCDS alignments of \emph{H. sapiens, P. troglodytes} and \emph{P. abelii} (references hg18, panTro2 and ponAbe2) were downloaded from the UCSC genome browser (\url{http://genome.ucsc.edu}). Only CCDS alignments satisfying the following requirements were retained for the subsequent analyses: divergence from human reference below $10\%$, no gene duplication in any species, start and stop codons conserved, no frame-shifting gaps, no gap longer than 30 bases, no nonsense codon, no gene shorter than 21 bases, no gene with different number of exons in different species, or genes in different chromosomes in different species (chromosomes 2a and 2b in non-humans were identified with human chromosome 2). From the remaining CCDSs (9,695 genes, 79,677 exons) we extracted synonymous sites. We only considered third co\-don positions where the first two nucleotides of the same co\-don were conserved in the alignment, as well as the first position of the next codon.

Furthermore, orang-utan SNP data for the two (Bor\-nean and Sumatran) populations considered, each with 5 sequenced individuals \cite{LockeEtAl11}, were kindly provided by X. Ma and are available online (\url{http://www.ncbi.nlm.nih.gov/projects/SNP/snp_viewTable.cgi?type=contact&handle=WUGSC_SNP&batch_id=1054968}). The final total number of synonymous sites included was 1,950,006.  Among them, a subset of 9750 4-fold degenerate synonymous sites that are polymorphic in the orang-utan populations were selected.

\section{Discussion}
\label{sec:discussion}

In this article, we propose an algorithm to approximate the maximum likelihood estimator in models with an intractable likelihood. Simulations and kernel density estimates of the likelihood are used to obtain the ascent directions in a stochastic approximation algorithm, so it is flexibly applicable to a wide variety of models. 

Alternative simulation based approximate maximum likelihood methods have been proposed that estimate the likelihood surface from samples from the whole parameter space that are obtained in an ABC like fashion \citep{CreelKristensen13, RubioJohansen13} or using MCMC \citep{Valpine04}. The maximum likelihood estimator is obtained subsequently by standard numerical optimization. Conversely, our method only uses simulations and estimates of the likelihood along the trajectory of the stochastic approximation algorithm, thereby approximating the maximum likelihood estimator more efficiently. This is along the lines of \citet{DiggleGratton84} where a stochastic version of the Nelder-Mead algorithm is used, but through the use of summary statistics we can drop the restrictive \textit{i.i.d.} assumption. In the setting of hidden Markov models, \citet{EhrlichEtAl13} have proposed a recursive maximum likelihood algorithm that also combines ABC me\-thod\-ology with the simultaneous perturbations algorithm. A thorough comparison of the properties of different approximate maximum likelihood methods is beyond the scope of this paper, but the three presented examples show that the algorithm provides fast and reliable approximations to the corresponding maximum likelihood estimators. 

The population genetics application where we estimate parameters of the evolutionary history of orang-utans demonstrates that very high-dimensional summary statistics (here: 56 dimensions) can be used successfully without any dimension-reduction techniques. Usually, high-dimensional kernel density estimation is not recommended because of the \textit{curse of dimensionality} \citep[e.g.,][]{WandJones}, but stochastic approximation algorithms are explicitly designed to cope with noisy measurements. To this end, we also introduce modifications of the algorithm that reduce the impact of single noisy likelihood estimates. In our experience, this is crucial in settings with a low signal-to-noise ratio.

Furthermore, the examples show that the AML algorithm performs well in problems with a high-di\-men\-sio\-nal and large parameter space: In the normal distribution example, the 10-dimensional maximum likelihood estimate is approximated very precisely even though the search space spans 200 times the standard error of $\hat{\Theta}_{\textrm{ML}}$ in each dimension.

However, we also observe a bias for a few of the estimated parameters. Partly, for example for one of the parameters of the queuing process, this can be attributed to a bias of the maximum likelihood estimator itself. In addition, it is known that the finite differences approximation to the gradient in the Kiefer-Wolfowitz algorithm causes a bias that vanishes only asymptotically \citep{Spall}, and that is possibly increased by the finite-sample bias of the kernel density estimator. In most cases though, the bias is smaller than the standard error of the approximate maximum likelihood estimator and can be made still smaller by carrying out longer runs of the algorithm. 

As sufficiently informative summary statistics are crucial for the reliability of both maximum likelihood and Bayesian estimates, the quality of the estimates obtained from our AML algorithm will also depend on an appropriate choice of summary statistics. This has been discussed extensively in the context of ABC \citep{FearnheadPrangle12, BlumEtAl13}. General results and algorithms to choose a small set of informative summary statistics should carry over to the AML algorithm. 

In addition to the point estimate, we suggest to obtain confidence intervals by parametric bootstrap. The bootstrap replicates can also be used for bias correction. Resampling in models where the data have a complex internal structure catches both the noise of the maximum likelihood estimator as well as the approximation error.  Alternatively, the AML algorithm may also complement the information obtained via ABC in a Bayesian framework: the location of the maximum a posteriori estimate can be obtained from the AML algorithm. 
 
The presented work shows the broad applicability of the approximate maximum likelihood algorithm and also its robustness in highly stochastic settings. With the implementation in the coalescent simulation program \texttt{msms} \citep{EwingHermisson10} its potential for population genetic applications can easily be explored further.

\section*{Acknowledgments}

Johanna Bertl was supported by the Vienna Graduate School of Population Genetics (Austrian Science Fund (FWF): W1225-B20) and worked on this project while employed at the Department of Statistics and Operations Research, University of Vienna, Austria. The computational results presented have partly been achieved using the Vienna Scientific Cluster (VSC). The orang-utan SNP data was kindly provided by X.~Ma. Parts of this article have been published in the PhD thesis of Johanna Bertl at the University of Vienna.

\bibliographystyle{abbrvnat_no}
\bibliography{PhD_all}

\end{document}